\documentclass[12pt]{article}
\usepackage{amsfonts}

\begin{document}

\title{Effect of the generalized uncertainty principle on Galilean and Lorentz transformations
}

\author{V. M. Tkachuk\\
Department for Theoretical Physics,\\ Ivan Franko National
University of Lviv,\\ 12 Drahomanov St., Lviv, UA-79005, Ukraine\\
e-mail: tkachuk@ktf.franko.lviv.ua\\
voltkachuk@gmail.com}

\maketitle

\begin{abstract}
Generalized Uncertainty Principle (GUP) was obtained in string theory and quantum gravity and suggested
the existence of a fundamental minimal length which, as was established,
can be obtained within the deformed Heisenberg algebra.
We use the deformed commutation relations or in classical case (studied in this paper) the deformed Poisson brackets, which are invariant with respect to the translation in configurational space.
We have found transformations relating coordinates and times of moving and rest
frames of reference in the space with GUP in the first order over parameter of deformation.
For the non-relativistic case we find the deformed Galilean transformation which is similar to the Lorentz one written for Euclidean space with signature $(+,+,+,+)$. The role of the speed of light here plays some velocity $u$
related to the parameter of deformation, which as we estimate is many order of magnitude larger than the speed of light
$u\simeq 1.2 \times 10^{22} c$.
The coordinates of the rest and moving frames of reference
for relativistic particle in the space with GUP satisfy the Lorentz transformation
with some effective speed of light.
We estimate that the relative deviation of this effective speed of light $\tilde c$ from $c$
is ${(\tilde c-c)/ c}\simeq 3.5\times 10^{-45}$.
The influence of GUP on the motion of particle and the Lorentz transformation in the first order over parameter of deformation is hidden in $1/c^2$ relativistic effects.
\end{abstract}

\section{Introduction}
The investigations in string theory and quantum gravity (see, e.g., \cite{gross, maggiore, witten})  lead to the
Generalized Uncertainty Principle (GUP)
\begin{eqnarray}
\Delta X\ge{\hbar\over2}\left({1\over \Delta P}+\beta\Delta P\right),
\end{eqnarray}
from which follows
the existence of the
fundamental minimal length $\Delta X_{\rm min}=\hbar\sqrt\beta$, which, as it is supposed, is
of order of Planck's length $l_p=\sqrt{\hbar G/c^3}\simeq 1.6\times 10^{-35}\rm m$.
A broad recent review on this subject can be found in paper \cite{Hos}.
We would like also to point out the recent discussion around the question whether
we can measure structures with precision better
than the Planck's length, which can be found in \cite{Hos12}.

It was established that minimal length
can be obtained in the frame of small quadratic modification (deformation) of the Heisenberg algebra \cite{Kem95,Kem96}
\begin{eqnarray} \label{XPdefq}
[X,P]=i\hbar(1+\beta P^2).
\end{eqnarray}
In the classical limit $\hbar\to 0$ the quantum-mechanical commutator for operators is replaced by the Poisson bracket for corresponding classical variables
\begin{eqnarray}
{1\over i\hbar}[X,P]\to\{X,P\},
\end{eqnarray}
which in the deformed case (\ref{XPdefq}) reads
\begin{eqnarray}\label{XP1}
\{X,P\}=(1+\beta P^2).
\end{eqnarray}

We would like to note that historically the first algebra of that kind in the relativistic case was proposed by Snyder in 1947 \cite{Snyder47}. But only investigations in string theory and quantum gravity renewed the interest in the studies of physical properties of classical and quantum systems in spaces with deformed algebras.
The observation that GUP can be obtained from the deformed Heisenberg algebra opens the possibility to study the influence of minimal length on properties of  physical systems on the quantum level as well as on the classical one.

Deformed commutation relations bring new difficulties in the quantum
mechanics as well as in the classical one.
There are known
only a few problems, which can be solved exactly.
Namely, one-dimensional harmonic
oscillator with minimal uncertainty in position \cite{Kem95} and
also with minimal uncertainty in position and momentum
\cite{Tkachuk1,Tkachuk2}, $D$-dimensional isotropic harmonic
oscillator \cite{chang, Dadic}, three-dimensional Dirac oscillator
\cite{quesne},
(1+1)-dimensional Dirac
oscillator within Lorentz-covariant deformed algebra \cite{Quesne10909},
one-dimensional Coulomb problem
\cite{fityo},
the
singular inverse square
potential with a minimal length \cite{Bou1,Bou2},
the (2+1) dimensional Dirac equation in a constant magnetic field
in the presence of a minimal length \cite{Men13}.
Three-dimensional
Coulomb problem with the deformed Heisenberg algebra was studied within the perturbation theory in \cite{Brau,Benczik},
where it was found that common perturbation theory does not work for $ns$-levels.
In
\cite{mykola,Stet,mykolaOrb} the modified perturbation theory was proposed, which allows to obtain
an explicit expression for corrections to $ns$-levels for hydrogen atom caused by the deformation of the Heisenberg algebra.
In \cite{Stet07} the scattering problem in the deformed space with minimal length was studied.
The ultra-cold
neutrons in gravitational field with minimal length were considered in
\cite{Bra06,Noz10,Ped11}.
The influence of minimal length on Lamb's shift, Landau levels, and tunneling current in scanning tunneling   microscope was studied in \cite{Das,Ali2011}.
The Casimir effect in a space with minimal length was examined in \cite{Frassino}.
In paper \cite{Vaki} the effect of noncommutativity and of the existence of a minimal length on the phase space of cosmological model was investigated.
The authors of paper \cite{Batt}
studied various physical consequences, which follow from the noncommutative Snyder space-time geometry.
The gauge invariancy in space with GUP was considered in \cite{Kob10}.
In paper \cite{Kempf12} the GUP and localization of a particle in a discrete space was studied.
Some consequences of the GUP-induced ultraviolet wave-vector cutoff in one-dimensional quantum mechanics was
studied in recent paper \cite{Sai13}.
The classical mechanics in a space with deformed Poisson brackets was studied
in \cite{BenczikCl,Fryd,Sil09}.
The composite quantum and classical system ($N$-particle system) in the deformed space with
minimal length was studied in \cite{Quesne10,Bui10}.

The study of deviation from standard quantum mechanics as well as from classical one caused by GUP gives
a possibility to estimate the upper bound for minimal length. The collection of upper boundes for minimal length
obtained form the investigation of different properties of different systems
can be found in recent paper \cite{Mar12}. The authors of this paper propose to use the
gravitational bar detectors to place an upper limit for a possible Planck-scale modifications
on the ground-state energy of an oscillator.
In \cite{Pik12} the authors propose to use the quantum-optical control
of the mechanical system to probe a possible deviation from the quantum commutation relation at the Planck
scale.

Note that deformation of the Heisenberg algebra and in classical case respectively Poisson brackets bring not only technical difficulties in solving of corresponding equations
but also bring problems of a fundamental nature.
One of them is the violation of the equivalence principle in
the space with minimal length \cite{Ali11}.
This is the result of assumption that the parameter of deformation
for
macroscopic bodies of different mass is unique.
In paper \cite{Quesne10} we showed that the center of mass of a macroscopic body in deformed space is
described by an effective parameter of deformation, which is essentially smaller than the parameters of deformation for particles constituting the body. Using the result of \cite{Quesne10} for the effective parameter of deformation in \cite{tkachuk12} we showed that the equivalence principle in the space with minimal length can be recovered.

In this paper we study the Galilean and Lorentz transformations in space with deformed Poisson brackets which correspond to the space with minimal length or GUP.
This paper organized as follows.
In section 2 starting from a non-relativistic Hamiltonian we find the Lagrangian of a particle in the space with deformed Poisson brackets.
In section 3 we study the invariancy of action with the Lagrangian obtained in section 2 and  find the deformed Galilean transformation for coordinates of a non-relativistic particle in one-dimensional space with GUP. In section 4 this result is generalized for the three-dimensional case. The Lorentz transformation for coordinates of relativistic particle in the space with GUP is studied in section 5. And finally, in section 6 we conclude the results.

\section{Hamiltonian and Lagrangian of a particle in deformed space}

In this section we find the Lagrangian of a classical particle in space with minimal length starting from the Hamiltonian formalism.
It is commonly supposed that Hamiltonian in deformed case has the form of Hamiltonian in non-deformed case where
instead of canonical variables of non-deformed phase space are written variables of deformed phase space.
So, the Hamiltonian of a particle (a macroscopic body which we consider as a point particle) of mass $m$ in the
potential $U(X)$ moving in one-dimensional configurational space reads
\begin{eqnarray}
H={P^2\over 2m} +U(X),
\end{eqnarray}
where $X$ and $P$ satisfy deformed Poisson bracket (\ref{XP1}).
This Poisson bracket allows the following coordinate representation
\begin{eqnarray}\label{RepXP1}
P={1\over\sqrt\beta}\tan({\sqrt\beta p}), \  \  X=x,
\end{eqnarray}
where small variables satisfy canonical Poisson bracket
\begin{eqnarray}\label{xp1}
\{x,p\}=1
\end{eqnarray}
and represent the non-deformed phase space.
The Hamiltonian in this representation reads
\begin{eqnarray}
H={\tan^2({\sqrt\beta p})\over 2m\beta} +U(x).
\end{eqnarray}
As we see, the deformation of the Poisson bracket in representation (\ref{RepXP1}) is equivalent to the deformation of kinetic energy.

We consider the linear approximation over the parameter of deformation $\beta$. In this approximation the Hamiltonian
reads
\begin{eqnarray}\label{Hr1}
H={p^2\over 2m} +{1\over 3}{\beta\over m}p^4 +U(x).
\end{eqnarray}
This Hamiltonian is similar to the relativistic one written in the first order over $1/ c^2$
\begin{eqnarray}
H_r=mc^2\sqrt{1+{p^2\over m^2c^2}}+U(x)=mc^2+{p^2\over 2m} -{1\over 8m^3c^2}p^4 +U(x) +O(1/c^4).
\end{eqnarray}
Introducing effective velocity
\begin{eqnarray}\label{cb}
 u^2={3\over 8\beta m^2}.
\end{eqnarray}
Hamiltonian (\ref{Hr1}) in the first order over $\beta$ or $1/u^2$
can be obtained from the following one
\begin{eqnarray}
H=-mu^2\sqrt{1-{p^2\over m^2 u^2}} +mu^2 +U(x).
\end{eqnarray}
This suggests that corrections to all properties related with deformations will be similar to relativistic ones
in the first order over $1/c^2$ but with an opposite sign before $1/c^2$. In particular it suggests that the
Galilean transformations in the first order over $\beta$ will be similar to the Lorentz one but with an opposite sign
before $1/c^2$. Let us show it subsequently.

Because $x$ and $p$ represent the non-deformed canonical space, the Lagrangian can be found in the traditional way
\begin{eqnarray} \label{defL}
L=\dot x p-H(x,p),
\end{eqnarray}
where $p$ is the function of $x$, $\dot x$ and can be found from equation
\begin{eqnarray}
\dot x={\partial H\over \partial p}= {p\over m}+{4\over 3}{\beta\over m}p^3.
\end{eqnarray}
In linear over $\beta$ approximation we find
\begin{eqnarray}
p=m\dot x\left(1-{4\over3}\beta m^2\dot x^2\right).
\end{eqnarray}

Substituting it into (\ref{defL}) we finally find the Lagrangian in the linear approximation over $\beta$
\begin{eqnarray}\label{Lcl}
L={m\dot x^2\over 2}-{1\over3}\beta m^3\dot x^4-U(x).
\end{eqnarray}
Similarly as Hamiltonian (\ref{Hr1}) this Lagrangian is very similar to the Lagrangian of a relativistic particle in first order over $1/c^2$, namely
\begin{eqnarray} \label{Lr}
L_r=-mc^2\sqrt{1-{\dot x^2\over c^2}}-U(x)=-mc^2 +{m\dot x^2\over 2}+{m\over8c^2}\dot x^4-U(x).
\end{eqnarray}
The difference is only in constant $mc^2$ and opposite sing in the last term.
Thus, we rewrite Lagrangian (\ref{Lcl})
as follows
\begin{eqnarray} \label{Ld1}
L=mu^2\sqrt{1+{\dot x^2\over u^2}}-mu^2-U(x),
\end{eqnarray}
where the effective velocity $u$ is the same as in (\ref{cb}). Of course,  Lagrangian (\ref{Ld1}) corresponds to
(\ref{Lcl}) only in the first order over $1/u^2$ or $\beta$.
The constant $-mu^2$ does not influence the equation of motion and can be omitted.

\section{Galilean transformation in deformed space}
To establish the Galilean transformation it is enough to consider free particle with Lagrangian
(\ref{Ld1}), where $U=0$. Omitting constant $-mu^2$ the Lagrangian for free particle in first order over $\beta$
reads
\begin{eqnarray}\label{Ld}
L=mu^2\sqrt{1+{\dot x^2\over u^2}}.
\end{eqnarray}
So, in the first order over parameter of deformation
$\beta$ the action reads
\begin{eqnarray} \label{Sd1}
S=mu^2\int_{t_1}^{t_2}\sqrt{1+{\dot x^2\over u^2}}dt=mu^2\int_{(1)}^{(2)}ds,
\end{eqnarray}
where
\begin{eqnarray}\label{int2}
ds^2= u^2 (dt)^2+(d x)^2
\end{eqnarray}
is squared interval in the Euclidean space whereas in relativistic case the second term has an opposite sign and space is pseudo-Euclidean.

Interval (\ref{int2}) is invariant under rotation in plane ($ut,x$). So, symmetry transformation reads
\begin{eqnarray}
x=x'\cos\phi+ut'\sin\phi, \\
ut=-x'\sin\phi+ut'\cos\phi.
\end{eqnarray}
The angle $\phi$ is related with the velocity $V$ of motion of the point $x'=0$ with respect to the rest frame of reference
\begin{eqnarray}
{V\over u}={x\over ut}= \tan\phi.
\end{eqnarray}
Then Galilean transformation reads
\begin{eqnarray}\label{GtrExact}
x={x'+Vt'\over \sqrt{1+V^2/u^2}}, \ \ t={t'-x'V/u^2\over\sqrt{1+V^2/u^2}}.
\end{eqnarray}
We call it the deformed Galilean transformation.
This transformation is very similar to the Lorenz one. The important difference is that here
we have an opposite sign before $1/u^2$ that is the result of positive $\beta$, for which just a minimal length exists.
For negative $\beta$ the minimal length is zero and according to (\ref{cb}) $1/u^2$ must be changed to
$-1/u^2$. In this case we have
common Lorentz transformations where instead of speed of light $c$ an effective velocity $u$ appears.
Note that
in fact this transformations are correct only in the first order over the parameter of deformation $\beta$,
which is related with $1/u^2$ [see (\ref{cb})]. So, in first order over parameter of deformation we find
\begin{eqnarray}\label{Gtr}
x=(x'+Vt')\left(1-{V^2\over2u^2}\right), \ \ t=t'\left(1-{V^2\over2u^2}\right)-x'{V\over u^2}.
\end{eqnarray}
In the limit $\beta\to 0$ or according to (\ref{cb}) $u \to \infty$ transformation (\ref{Gtr}) recover
ordinary Galilean transformation. Here it is interesting to note that transformation (\ref{GtrExact}) or (\ref{Gtr}) is one of the possible
transformations, which can be obtained in the frame of the following question asked in Special Relativity Theory:
what the most general transformations
of spacetime were that implemented the relativity principle, without
making use of the requirement of the constancy of the speed of light?
For details, see section ``Algebraic and Geometric Structures in Special Relativity'' in review
\cite{SR}.

Here it is worth to mention the result of paper \cite{tkachuk12} where we showed that for a body of mass $m$
the parameter of deformation reads
\begin{eqnarray}\label{Effbeta}
\beta={\gamma^2\over m^2},
\end{eqnarray}
where $\gamma$ is the same constant for bodies of different mass. It is interesting to note that constant $c\gamma$ is dimensional.
Stress that that only the relation (\ref{Effbeta}) as was showed
in paper \cite{tkachuk12}
leads to recovering of the equivalence principle in the deformed case.
As a result of (\ref{Effbeta}) we have
\begin{eqnarray}\label{cbg}
u^2={3\over 8\gamma^2}.
\end{eqnarray}
and thus the effective velocity does not depend on mass of a body. It means that Galilean transformation
is the same for coordinates of particles of different mass as everybody feels it must be.

\section{Three-dimensional case}
The generalization of obtained Galilean transformation on three dimensional case is straightforward.
We consider deformed  algebra, which is invariant with respect to translations in configurational space.
Different algebras of this type can be found in \cite{Fryd} (see also references therein). One of the possible algebra of this type reads
\begin{eqnarray}\label{XPdef}
[X_i,P_j]=i\hbar\sqrt{1+\beta P^2}\left(\delta_{i,j}+\beta P_i P_j\right),\\
\label{XPdef2}
{}[X_i,X_j]=[P_i,P_j]=0.
\end{eqnarray}
and can be obtained using the representation
\begin{eqnarray}\label{rep3D}
X_i=x_i, \ \ P_i={p_i\over\sqrt{1-\beta p^2}},
\end{eqnarray}
where
${\bf x}=(x_1,x_2,x_3)$, ${\bf p}=(p_1,p_2,p_3)$ represent the coordinates and momentum in non-deformed space
with canonical commutation relations
\begin{eqnarray}
[x_i,p_j]=\hbar\delta_{i,j},\ \ [x_i,x_j]=[p_i,p_j]=0.
\end{eqnarray}
Note that in the momentum representation as follows from (\ref{rep3D}) $p^2<1/\beta$ and as a result there is nonzero minimal uncertainty in position or minimal length.
The algebra given by (\ref{XPdef}) and (\ref{XPdef2}) is invariant with respect to the transformation $\bf X=\bf X'+\bf a$ and thus is translation-invariant in configurational space. It means that the space is uniform.

Now we consider the classical limit $\hbar\to 0$. Then the deformed Poisson brackets corresponding to algebra (\ref{XPdef}), (\ref{XPdef2}) read
\begin{eqnarray}\label{XPdefP}
\{X_i,P_j\}=\sqrt{1+\beta P^2}\left(\delta_{i,j}+\beta P_i P_j\right),\\
{}\{X_i,X_j\}=\{P_i,P_j\}=0.
\end{eqnarray}
The Hamiltonian in representation (\ref{rep3D}) is the following
\begin{eqnarray}\label{H3}
H={1\over2m}{p^2\over 1-\beta p^2}+U({\bf x})={p^2\over 2m}+{\beta\over 2m}p^4 +U({\bf x})+O(\beta^2),
\end{eqnarray}
where in our consideration we restrict oneself up to
to the first order over $\beta$.

Similarly as in one-dimensional case we find the Lagrangian corresponding to Hamiltonian (\ref{H3}). First, we find
the relation between the velocity and momentum of the particle
\begin{eqnarray}
\dot x_i={1\over m}{p_i\over (1-\beta p^2)^4}={p_i\over m}(1+2\beta p^2) + O(\beta^2)
\end{eqnarray}
and in first order over $\beta$ we obtain
\begin{eqnarray}
p_i=m\dot x_i(1-2\beta \dot x^2).
\end{eqnarray}
The Lagrangian in this approximation reads
\begin{eqnarray}
L={m\dot {\bf x}^2\over 2}-{\beta m^3\over 2}\dot {\bf x}^4-U({\bf x}).
\end{eqnarray}
Similarly as in one-dimensional case this Lagrangian in the first order over $\beta$ can be written in the form (\ref{Ld1}) and the action of free particle with $U=0$ for three dimensional case takes form (\ref{Sd1}) where
\begin{eqnarray}\label{int4}
ds^2=u^2 (dt)^2+(d x_1)^2+(d x_2)^2+(d x_3)^2,
\end{eqnarray}
here
\begin{eqnarray}\label{cb3}
u^2={1\over 4\beta m^2}={1\over 4\gamma^2}.
\end{eqnarray}

In paper \cite{tkachuk12} from the suggestion that minimal length for electron is
of order of Planck's length we estimate $\gamma$.
Doing similarly we suggest that for  electron $\hbar \sqrt\beta=l_p$. Then taking into account relation
(\ref{Effbeta}) and substituting for $m$ the mass of electron we find
$c\gamma\simeq 4.2\times 10^{-23}$ that reproduce the result of paper \cite{tkachuk12}.
Using this result we find that
$u\simeq 1.2 \times 10^{22} c$ which is many order of magnitude large than the speed of light.

Thus, when the second frame of reference $(t',{\bf x'})$ moves with respect to the first one $(t,{\bf x})$ with velocity
$V$ along axis $x_1$ then Galilean transformation of coordinate $x'_1$ and time $t'$ to $x_1$ and time $t$ satisfies (\ref{Gtr}), other coordinates are not changed $x_2=x'_2,\ \ x_3=x'_3 $.

\section{Lorentz transformation in deformed space}

In this section we generalize the above consideration for the relativistic case.
Let us start from the one-dimensional relativistic Hamiltonian for free particle
\begin{eqnarray}
H=mc^2\sqrt{1+{P^2\over m^2c^2}},
\end{eqnarray}
where position and momentum satisfy deformed Poisson bracket (\ref{XP1}).
Using representation (\ref{RepXP1}) in the first order over $\beta$ and $1/ c^2$ this Hamiltonian reads
\begin{eqnarray}
H=m^2c^2+{p^2\over 2m}-\left({1\over 8m^2c^2}-{\beta\over 3}\right){p^4\over m}.
\end{eqnarray}
Introducing notation
\begin{eqnarray}\label{Effc1}
{1\over 8m^2\tilde c^2}={1\over 8m^2c^2}-{\beta\over 3}
\end{eqnarray}
we find that this Hamiltonian can be obtained in first order over $1/\tilde c^2$ from the following one
\begin{eqnarray}\label{Hdef}
H=m\tilde c^2\sqrt{1+{p^2\over m^2\tilde c^2}} -m\tilde c^2 +mc^2.
\end{eqnarray}
We suppose that $\beta$ is much smaller than $1/m^2c^2$. Then this Hamiltonian corresponds to the relativistic one
but with an effective velocity $\tilde c$, which is defined by (\ref{Effc1}). Note that $\tilde c> c$ and
$\tilde c\to c$ when $\beta\to 0$. Thus transformation relating coordinates and time of two reference frames
is the Lorentz transformation which contains instead of speed of light $c$ the effective speed $\tilde c$.

Taking into account (\ref{Effbeta})
we find that (\ref{Effc1}) reads
\begin{eqnarray}\label{Effc1D}
{1\over \tilde c^2}={1\over c^2}-{8\over 3}\gamma^2
\end{eqnarray}
and thus the effective speed of light does not depend on the mass of a body. It means that the Lorentz transformation
is the same for particles of different mass as it must be.

The generalization on three-dimensional case is straitforward. For the deformed algebra given by
(\ref{XPdef}), (\ref{XPdef2}) we obtain the Hamiltonian
in form (\ref{Hdef}) where effective velocity is defined by
\begin{eqnarray}\label{Effc3D}
{1\over \tilde c^2}={1\over c^2}-4\gamma^2.
\end{eqnarray}
So, similarly as in the one-dimensional case
the Lorentz transformation
contains instead of speed of light the effective speed of light.
Note that for different deformed algebras we obtain the same result, only the factor before
$\gamma^2$ will be different.
In general we can write
\begin{eqnarray}\label{EffcG}
{1\over \tilde c^2}={1\over c^2}-{1\over u^2},
\end{eqnarray}
where $u=\alpha c/\gamma$ and $\alpha$ is a multiplier different for different algebras.
The relative deviation of the effective speed of light $\tilde c$ from $c$ in the first order over the parameter of deformation
$\beta$ or $\gamma$ reads
\begin{eqnarray}
{\tilde c-c\over c}=2c^2\gamma^2\simeq 3.5\times 10^{-45},
\end{eqnarray}
here we use that $c\gamma\simeq 4.2\times 10^{-23}$ [see explanation after eq. (40)].

\section{Conclusions}

In the present paper we have found the transformations relating coordinates and times of particle in moving and rest
frames of reference in the space with GUP or minimal length in the first order over the parameter of deformation.
For the description of the space with GUP we used the deformed algebra which is invariant with respect to translation in the configurational space. In the classical case considered in this paper we have corresponding deformed Poisson brackets.

For the non-relativistic case we find that this transformation is similar to the Lorentz one but for space with signature
$(+,+,+,+)$. We call it the deformed Galilean transformation and it is rotation in Euclidian space. The role of the speed of light here plays some  velocity
$u$ which is inverse to ${\sqrt\beta} m$.
It is important to note that, as we shown in our previous paper \cite{tkachuk12}, the equivalence principle and independence of kinetic energy on composition of a body require that ${\sqrt\beta} m=\gamma$ is constant and does not depend on the mass of the body.
Doing similarly as in paper \cite{tkachuk12} we suggest that minimal length for electron is of order of Planck's length and set $\hbar \sqrt\beta=l_p$, then $c\gamma\simeq 4.2\times 10^{-23}$.
Applying this result to the deformed Galilean transformation we find that this transformation is the same for bodies of different mass as everybody feels it must be and also estimate the effective velocity which is
many orders of magnitude larger than the speed of light $u\simeq 1.2 \times 10^{22} c$.
Therefore, the effect of GUP on the motion of particle and
Galilean transformation is much order smaller the relativistic one.
Note that the deformed Galilean transformation in contrary to ordinal one contains also the transformation of time which thus is not absolute in
space with GUP.
In the limit $\beta\to 0$ or $\gamma\to 0$ the deformed Galilean transformation recovers
the ordinary one.
Let us now explain qualitatively why the deformed Galilean transformation is similar to the Lorentz one.
Considering non-relativistic case we start from the common non-relativistic Hamiltonian written in deformed variables.
But using the representation of deformed variables over non-deformed ones we find that the Hamiltonian in the first order over the parameter of deformation contains an additional term proportional to $p^4$.
This Hamiltonian is similar to the relativistic one written in the first order over $1/c^2$ but with an opposite sign before $p^4$. This very sign leads to a four-dimensional Euclidean space with signature $(+,+,+,+)$ in contrary to the ordinary relativistic case with pseudo-Euclidean space with signature $(+,-,-,-)$.
It is interesting to note that deformed Galilean transformation obtained here for space with GUP is Euclidian rotation and it is one of the possible
transformations, which can be obtained in the frame of the following question asked in Special Relativity Theory:
what the most general transformations
of spacetime were that implemented the relativity principle, without
making use of the requirement of the constancy of the speed of light?
For details, see section ``Algebraic and Geometric Structures in Special Relativity'' in review
\cite{SR}.

The similarity of the deformed Galilean transformation to the Lorentz one forced us to study the relativistic particle in a space with GUP predicting that the effect of GUP can be hidden in the relativistic effect.
We describe the relativistic particle in space with minimal length or GUP by the relativistic Hamiltonian which contains deformed variables instead of non-deformed ones.
Using the representation of deformed variables over non-deformed ones we find that the Hamiltonian in the first order over the parameter of deformation and first order over $1/c^2$ has also a relativistic form in non-deformed variables with some effective speed
of light $\tilde c$. Therefore, coordinates of a relativistic particle in the rest and moving frames of reference
in space with minimal length satisfy the Lorentz transformation
with an effective speed of light.
Similarly as in the non-relativistic case the effective speed of light does not contain the mass of particle when
condition ${\sqrt\beta} m=\gamma$ holds and thus in this case the Lorentz transformation is the same
for coordinates and time of particles of different masses.
We estimate that the relative deviation of effective speed of light $\tilde c$ from $c$
is very small ${(\tilde c-c)/ c}\simeq 3.5\times 10^{-45}$.
Finally let us note that the influence of GUP on the motion of particle and the Lorentz transformation in the first order over the parameter of deformation is hidden in $1/c^2$ relativistic effects.
\section*{Acknowledgment}

I am grateful to Dr. T. Mas{\l}owski for drawing my attention to review \cite{SR}.

\end{document}